
\font\titlefont = cmr10 scaled \magstep2
\magnification=\magstep1
\vsize=20truecm
\voffset=1.75truecm
\hsize=14truecm
\hoffset=1.75truecm
\baselineskip=20pt

\settabs 18 \columns

\def\b{\bigskip}
\def\bb{\bigskip\bigskip}

\def\ce{\centerline}

\def\no{\noindent}




\rightline{UMDHEP-93-201}
\rightline{ISJ-5079}
\rightline{May 1993}

\ce{\titlefont { Effective Lagrangian Approach to Electroweak}}
\ce{\titlefont { Baryogenesis:~~higgs mass limit}}
\ce{\titlefont{ and electric dipole moments of fermion}}

\b
\ce{ X.~ Zhang}
\ce{
Department of Physics \& Astronomy, University
 of Maryland,}
\ce{ College Park, MD 20742}

\ce{and}

\ce{B.-L. ~Young}
\ce{Department of Physics \& Astronomy, Iowa State University,}
\ce{ Ames, Iowa
 50011}

\b

\ce{\bf ABSTRACT}
\b
A natural solution to the hierarchy problem of the
standard model
is to assume new physics to
appear
at the TeV scale. We parametrize the effects of this new physics in terms of
effective lagrangian and examine its impacts on electroweak baryogenesis.
We point out that
with such an effective lagrangian
successful electroweak baryogenesis implies: i)
Higgs boson lies within the reach of LEP II; ii) electric dipole moments of
electron and neutron are detectable in the near future.

\bb

\bb
\filbreak

Recent experimental data from LEP and SLC has provided remarkable checks
of the validity of the standard model[1]. The data is in agreement with
the theoretical
prediction of the model to a level which is  better than one percent[2].
Even though
the standard model is perfectly consistent with all evidence gathered
to date, it raises as many questions as it answers. The so-called
hierarchy problem is one of them, and it is based on the observation that
the quadratic divergence of the Higgs sector makes it difficult
to explain a widely separated hierarchy between the Fermi scale and a very
high scale of new physics,
$\Lambda$. Obviously, the hierarchy problem is solved if $\Lambda$ is
also around the
Fermi scale. In this case, one would expect new physics
beyond the standard model to appear at the TeV scale.

Another indication for the possible existence of new physics is from
the current understanding of the baryon
asymmetry of the universe. In Sakharov's original
proposal[3], three ingredients are required:
 a) baryon number violation; b) CP and C violation; c)
thermodynamic nonequilibrium. Indeed the three conditions
can be satisfied[4] in the standard model. Unfortunately the net matter-
antimatter asymmetry generated in the standard model is too small to
yield anything
like the observed asymmetry.
The effect of the CP violation in the standard model from the
Kobayashi-Maskawa (KM) phase
is much too small\footnote{[F.1]}{There are some interesting proposal
by Shaposhnikov and his collobarator to enhance the CP violation in the
standard model at high temperature[5].}.
In addition, in order to protect
the asymmetry so that it can survive until the present, the Higgs mass
cannot be heavier than O(40 GeV). This limit lies below the
present experimental lower bound $m_H \geq~ 60 ~GeV$.
 By these considerations, baryogenesis has an appealing
implication of new physics beyond the standard model.

Assuming the existence of new physics at the TeV scale,
we examine in this paper its impacts on
electroweak baryogenesis\footnote{[F.2]}{
Other scenarios for baryogenesis
can be found in Refs.[6, 8]. } in terms of an effective lagrangian.

As the structure of the underlying theory beyond the standard model
is presently unknown, it is useful to
parametrize the
effects of
 new physics in terms of effective lagrangian. In the literature,
two schemes based on linear and nonlinear realization of $SU_L(2) \times
U_Y(1)$ are used to construct the effective lagrangian.
The nonlinear
scheme, which puts less constraints on the effective lagrangian than the
linear scheme does, allows more operators and provides
more information for experiment to test. However, the effective lagrangian
in the nonlinear scheme is built on the broken phase of $SU_L(2) \times
U_Y(1)$, so it is not useful for baryogenesis, which need to consider the
change of vacuum expectation value of the Higgs field with temperature.
We will restrict ourselves to the linear scheme.

In the linear scheme, the effects of new physics are described by
higher-dimension operators $O^i$,

$${
{\cal L}^{new} =\Sigma_{i} {c_{i}\over \Lambda^{{d_i}-4}}O^i ~~,     }
\eqno(1)$$
 \no where $d_i$ are integers greater than 4.
The operators $O^i$ which have dimension ${[mass]}^{d_i}$ are
$SU_L(2)\times U_Y(1)$ gauge invariant and contain
only the
 standard model fields.
The parameters $c_i$,
determining the strength of the contribution of operator $O^i$, can
in principle be calculated
by matching the effective theory
with the underlying theory.
However, they are taken as free parameters here since we don't know the
underlying theory.

First of all, let us comment on the Higgs mass
limit in this model. In a recent paper[7], it has been demonstrated
that in the effective theory the Higgs mass can be relaxed to the
present experimentally allowed region. The arguments is straightforward:
in the effective lagrangian, higher dimension
operators change the relation between the Higgs mass and the quartic coupling
of the Higgs field. Consequently, the requirement of avoiding washout
of the baryon asymmetry is described as followes,
$${
m_H^2 < {( 40~GeV)}^2 ~ +~ 8 {v^4 \over \Lambda^2},
}\eqno(2)$$

\no where the number in bracket corresponds to the approximate
Higgs mass in the limit of $\Lambda \rightarrow \infty$;
and $v
\sim 250 ~GeV$ is the vacuum expectation value of Higgs field.
Equation (2) implies the $\Lambda \leq~ O(4{\rm TeV})$ .
 Certainly, the high energy
scale of $\Lambda \sim$1 TeV, which is required to solve
the hierarchy problem, allows an Higgs mass
which satisfies comfortably
the experimental lower bound of 60 GeV.
Thus an indication of electroweak baryogenesis is that Higgs may be discovered
at LEP II.

Having solved the ``Higgs mass limit" problem we now consider
the impacts of new CP violation required to produce enough baryon asymmetry
on
 the electric dipole moments of electron and neutron.
 Firstly, we give a brief summary of electroweak baryogenesis[8].

The baryon asymmetry is related to the total change of the
Chern-Simon number during the electroweak phase transition:

$${
{\alpha_W \over {4\pi}} \int^{\infty}_{t_c} d^4 x W_{\mu\nu} {\tilde
W}^{\mu\nu}
{}~~,}\eqno(3)$$

\no where
${\tilde W}^{\mu\nu}
={1\over 2}{\epsilon}^{\mu\nu\lambda\rho}{W_{\lambda\rho}}$, and
$t_c$ corresponds to the time when
electroweak phase transition startes.

To have a net change of the Chern-Simon number,
 one needs a term breaking P and CP.
The simplest operator is\footnote{[F.3]}{This operator
has been considered in Ref[9]. Here we use an equivalent, but slightly
 different, method to derive baryon asymmetry.}
$${
O_w= c_w
{g^{2}\over {8\pi^{2}}}
{\phi^{2}\over \Lambda^{2}} TrW_{\mu\nu}{\tilde W}^{\mu\nu} ,}\eqno(4)$$

\no where $\phi$ is
the neutral component of the standard model Higgs doublet,
$c_w$ a free parameter.
Simply one can see that
 this operator induces
 a ``chemical potential" $\mu = {d\over dt} (
{c_w {\phi^{2}\over \Lambda^{2}} } ) $ for
the Chern-Simon number,
with which the sphaleron process is biased.
We have

$${ \eqalign{
{ n_B }  &= 3 \int {{\Gamma_{sph}}\over T}~\mu~ dt \cr
    &=  3 \int {{ \Gamma_{sph}}\over T}~ d(c_w{\phi^{2}\over
\Lambda^{2}}) , \cr} } \eqno(5)$$
\no where the factor 3 is the number of generations, $\Gamma_{sph}$ the
sphaleron rate.

 During electroweak phase
transition, Higgs field rolls from false vacuum $(\phi = 0)$ down to the
true vacuum $( \phi \not= 0 )$. Through CP violating operator $O_w$
the sphaleron process is biased, resulting in
net baryon asymmetry. The final asymmetry can be obtained by integrating
Eq.(5). However, $\Gamma_{sph}$ in the integral is not well known.
Instead, what we know about $\Gamma_{sph}$ are:

\item{i)}  In the unbroken phase
$${
\Gamma_{sph} = \kappa {(\alpha_{W} T)}^{4}, }\eqno(6)$$
\no
where $\kappa$ subsumes our ignorance about the exact rate, which is
taken to be $0.1~ \sim ~ 1$ from numerical simulation[10].

\item{ii)} In the broken phase
$${
\Gamma_{sph} = \gamma {(\alpha_{W} T)}^{-3} M_W^{7} e^{-{E_{sph}\over T}}
, }\eqno(7)$$
\no where $E_{sph} \sim {g \phi(T) \over \alpha_{W}}$ and
$\gamma$ a constant evaluated numerically in Ref.[11].

\item{iii)} $\Gamma_{sph}$ in Eq.(5) will be much suppressed when the
Higgs vacuum expectation value exceeds the critical value $\phi_c(T)$.
In Ref.[12], it is estimated that
 $\phi_{c}(T) \sim 14 {\alpha_{W}T\over g}$. There are also some arguments
for smaller $\phi_c(T)$[13].

\b
\no So practically,
 the integral can be estimated by the following approximation:
  inserting a step function $\Theta ( \phi_{c} - \phi )$ in the integrant
and
taking $\Gamma_{sph}$ to be
$\kappa {(\alpha_{W} T)}^{4}$.
Now we arrive at the final
  ratio of baryon number density to entropy:

$${\eqalign{
{n_B \over s}& =
                 { 3~\times 45~\times 0.5  \over {2 \pi^{2} g_{*}}}
                \kappa ~c_w~ {\alpha}^{4}_{W}{({T\over \Lambda})}^2\cr
 & \sim  4 \times 10^{-2} \kappa ~c_w~ {\alpha_{W}^{4}}
{( {T\over \Lambda}) }^{2} , \cr }  }\eqno(8)$$
\no where $g_{*} \sim O(100)$ is the statistical weight factor appearing
in the formula of entropy $s$
and $T \sim O(100~ {\rm GeV})$ for weak phase transition.
Taking $\Lambda \sim 1 TeV$, one can see that to
 explain the experimental value
$${
{n_B \over s} \sim (0.4 - 1.4) \times 10^{-10}, }\eqno(9)$$
it is required that for $\kappa \sim 1-0.1$
$${
    c_w \geq 0.1-1 ~~~ , }\eqno(10)$$

\no which is not an unreasonable value to expect.

Let us calculate the electric dipole moments of electron and neutron induced
by $O_w$.
 At zero temperature,
 $\phi = {1\over {\sqrt 2} }{\pmatrix{ 0 \cr  v+ H \cr} } $,
with H being the standard Higgs particle. After diagonalizing the mass
matrix of $W^{3}_\mu~{\rm and}
 ~ B_\mu$, one can see that an effective Higgs-photon-photon
vertex is induced by Eq. (4):
$${
{\cal L}_{H \gamma \gamma}\sim c_w {\alpha_{em}\over {32\pi}} {H\over v} {
F_{\mu\nu}{\tilde F}^{\mu\nu}~} , }\eqno(11)$$

\no where $F_{\mu\nu}$ is the field strength of $U_{em}(1)$. In Eq.(11)
 we have taken $\Lambda \sim 1 TeV$. This operator
breaks CP and contributes to electric dipole moments of the fermion, which
can been seen by attaching the Higgs and one photon to the fermion line[14].
 We use dimensional regularization to
regularize
this divergent
 integration. Since the fundamental theory is assumed to be renormalizable, the
infinities should be absorbed in the counterterms appearing in the effective
lagrangian
$${
   {m_\psi \over \Lambda^3} i {\overline{\Psi_L}}\phi \sigma_{\mu\nu}
          \Psi_R F^{\mu\nu} + h.c ~~, }
\eqno(12)$$
\no where $m_\psi$ is the mass of fermion field $\Psi$.

The inclusion of such counterterm would introduce a free parameter as to
preclude predictive power. Instead, what has been done in the literature
is to focus on certain ``Log-enhanced" term. Practically, we compute
the leading correction of operator $O_w$
to the electric dipole moments of fermion in
$\overline{MS}$ scheme, setting the counterterm in (12) to zero for
$\mu$, the renormalization scale, to be $\Lambda$. We have now
$${
{ d_{\psi}\over e}
 \sim Q_{\psi}{ c_w \alpha_{em} m_{\psi} \over {32 \pi v^2}}
{1\over { ( {4 \pi})}^2 } \{ {1\over 2} + 2\int_0^1 dx
 \int_0^{1-x} dy \ln {\Lambda^2 \over {m_H^2 y + x^2 m_{\psi}^2} } \}
,}\eqno(13)$$
\no where $Q_{\psi}$ is the electric charge of field $\Psi$.
In this formula, $c_w$ is fixed as shown in Eq.(10)
 and
$m_{H}$ is bounded from above as we mentioned before.
 Thus we have for electron and neutron
$${
 {d_{e}\over e} \ge 10^{-28}-10^{-27} cm ,}\eqno(14.a)$$

$${
 {d_{n}\over e} \ge 10^{-27}-10^{-26} cm .}\eqno(14.b)$$

\no
 In
the derivation of the electric dipole moment of neutron we have used
nonrelativistic
quark model, where $d_{n}\sim d_{d}, ~d_{u}$. Clearly the predicted values
in Eqs.(14) are very close to the present experimental limits.

In conclusion,
we have demonstrated that a solution to the hierarchy problem
of the Higgs sector of the standard model opens the possibility
for electroweak baryogenesis. We have predicted the electric
dipole moments of electron and neutron from
the new CP violation term required to produce enough
baryon asymmetry of the universe.
We should point out that predictions of the electric dipole moments from
baryogenesis also exist in two
Higgs model[15], minimal supersymmetric standard model[16] and left-right
symmetric model[17].
However, the effective lagrangian approach to electroweak
baryogenesis used in this article
depends only on the general form of operators available at low energies, and
can describe theories in which no exotic particles
appear at low energies.
Examples of this kind of theory
include certain ``composite Higgs" models[18] and composite models[19].

 One may wonder why we choose among many possible operators only the
particular one given in Eq. (4)
for baryogenesis.
If one assumes that the new physics
couples very weakly to quarks and leptons, and
 contributes mainly to the gauge bosons' propagator corrections (the
so-called
 ``oblique corrections")[20], there are only two operators
of the lowest dimension which are relevent to
electroweak baryogenesis:
$${
O_w \sim {\phi^2 \over \Lambda^2} Tr W_{\mu\nu} {\tilde W}^{\mu\nu}~~,
}\eqno(15)$$

$${
O_g \sim {\phi^2 \over \Lambda^2} TrG_{\mu\nu} {\tilde G}^{\mu\nu}~~,
}\eqno(16)$$
\no where $G_{\mu\nu}$ is the gluon field strength. The
$O_g$ is important only in the case of a thick enough bubble wall so that
QCD anomalous interaction is in thermal equilibrium[21].

To conclude, we would like to point out that
the electroweak baryogenesis calculations available so far are
qualitative and the quatitative results obtained are probably
only accurate to within
a couple of order of magnitude.
 Thus future
low energy measurements on Higgs mass, $d_e$
and $d_n$ will test the present knowledge about electroweak baryogenesis.

\bb
X.Z. would like to thank R.N. Mohapatra and R.D. Peccei for helpful
discussions.
The work of XZ is supported by a grant from National Science Foundation.
BLY is suppport in part by the U.S. Department of Energy under contract
No. W-7405-Eng-82, Office of Energy Research (KA-01-01), and Division of
High Energy and Nuclear Physics.

\bb

\ce{\bf References}
\b
\item{[1]}S. Weinberg, Phys. Rev. Lett. $\bf 19$, (1967) 1264;
    A. Salam, {\it in} Elementary Particle theory: relativistic
groups and analyticity, Nobel Symp. No. 8, ed. N. Svartholm
(Almqvist and Wiksell, Stockholm, 1969) p. 367;
S.L. Glashow, J. Iliopoulos and L. Maiani, Phys. Rev. $\bf D2$,
(1970) 1285.

\item{[2]}
P. Langacker, UPR-0555T, March 24, 1993.

\item{[3]}A. Sakharov, JETP lett. 5 (1967) 24.

\item{[4]}V. Kuzmin, V. Rubakov and M. Shaposhnikov, Phys. Lett.
$\bf B155$, (1985) 36.

\item{[5]}
M. Shaposhnikov, Phys. Lett. $\bf B277$ (1992) 324; Erratum, Phys. Lett.
$\bf B282$, (1992) 483;
G.R. Farrar and M.E. Shaposhnikov, Phys. Rev. Lett.
$\bf 70$, (1993) 2833.

\item{[6]}
      A. Dolgov, Phys. Rept. $\bf 222$ (1990) 222.

\item{[7]}X. Zhang, Phys. Rev. $\bf D47$, (1993) 3065.

\item{[8]}A. Cohen, D. Kaplan and A. Nelson,
UCSD-PTH-93-02, BUHEP-93-4, January 1993 and references therein.

\item{[9]}
 M. Dine, P. Huet, R. Singleton and L. Susskind,
Phys. Lett. $\bf B257$, (1991) 351.

\item{[10]} J. Ambjorn, T. Askgaad, H. Porter and
M. Shaposhnikov, Phys. Lett. $\bf B244$, (1990) 479.

\item{[11]}L. Carson, X. Li, L. McLerran and R. Wang,
          Phys. Rev. $\bf D42$ (1990) 2127.

\item{[12]}B. Liu, L. McLerran, N. Turok, Phys. Rev. $\bf D46$,
(1992) 2668.

\item{[13]}M. Dine, P. Huet and R. Singleton,
Nucl. Phys. $\bf B375$ (1992) 625.

\item{[14]}S. Barr and A. Zee, Phys. Rev. Lett. $\bf 65$, (1990) 21.

\item{[15]}A. Kazarian, S. Kuzmin and M. Shaposhnikov,
           Phys. Lett. $\bf B276$, (1992) 131.

\item{[16]}A. Cohen and A. Nelson, Phys. Lett. $\bf B279$, (1992) 111.

\item{[17]}R. Mohapatra and X. Zhang, Phys. Rev. $\bf D46$, (1992) 5331.

\item{[18]} H. Georgi, {\it {Lectures at les Houches Summer School}}
(1985) and references therein.

\item{[19]}R.D. Peccei,
in the proceeding of
{\it{ The 1987 Lake Louise Winter Institute, Selected Topics in Electroweak
Interaction}}, J.M. Cameron et al., ed (World Scientific, 1987)
and references therein.

\item{[20]} B.W. Lynn, M.E. Peskin and R.G. Stuart, in {\it {Physics
at LEP}}, ed. J. Ellis and R. Peccei, CERN 86-02 (1986)I, 90;
D.C. Kennedy and B.W. Lynn, Nucl. Phys. $\bf B332$, (1989) 1;
M.E. Peskin and T. Takeuchi, Phys. Rev. Lett. $\bf 65$ (1990) 964.

\item{[21]}R. Mohapatra and X. Zhang, Phys. Rev. $\bf D45$, (1992) 2699.

\bye